\newif\ifpdf
\ifx\pdfoutput\undefined
  \pdffalse              
\else
  \pdfoutput=1           
  \pdftrue
\fi
\documentclass[aps,prb,twocolumn,superscriptaddress]{revtex4}
\ifpdf
  \usepackage[pdftex,
        colorlinks=true,
        pdftitle={A New face on old code},
        pdfauthor={P.F. Peterson, Th. Proffen, R.L. Mikkelson,
T. Kozlowski, D.J. Mikkelson, G. Cooper, and T.G. Worlton},
        pdfsubject={},
        pdfkeywords={NOBUGS2002/024},
        pdfpagemode=None,
        bookmarksopen=true]{hyperref}
  \usepackage{thumbpdf}
\fi
\usepackage{graphicx}
\usepackage{amssymb}

\begin{document}


\title{Sharing data between facilities: using the NeXus time-of-flight\\
powder diffractometer file format - NOBUGS2002/036}


\author{P.F. Peterson}
\affiliation{Intense Pulsed Neutron Source, Building 360, 9700 South
Cass Ave, Argonne, IL 60439-4814, USA}
\author{Th. Proffen}
\affiliation{Los Alamos National Laboratory, LANSCE-12, Mailstop H805,
Los Alamos, NM 87544, USA}
\author{R.L. Mikkelson}
\affiliation{Department of Mathematics, Statistics and Computer
Science, University of Wisconsin-Stout, Menomonie, WI 54751, USA}
\author{T. Kozlowski}
\affiliation{Los Alamos National Laboratory, LANSCE-12, Mailstop H805,
Los Alamos, NM 87544, USA}
\author{D.J. Mikkelson}
\affiliation{Department of Mathematics, Statistics and Computer
Science, University of Wisconsin-Stout, Menomonie, WI 54751, USA}
\author{G. Cooper}
\affiliation{Los Alamos National Laboratory, LANSCE-12, Mailstop H805,
Los Alamos, NM 87544, USA}
\author{T.G. Worlton}
\affiliation{Intense Pulsed Neutron Source, Building 360, 9700 South
Cass Ave, Argonne, IL 60439-4814, USA}


\date{\today}

\begin{abstract}
NeXus is an international standard data format intended to reduce the
need for redundant software development efforts in the neutron and
x-ray scattering communities. As the NeXus standard matures it is
starting to be used at laboratories for storing raw data. The Manuel
Lujan Jr. Neutron Scattering Center (MLNSC) at Los Alamos National
Laboratory and the Intense Pulsed Neutron Source (IPNS) at Argonne
National Laboratory have been working with NeXus in an effort to share
data and software. MLNSC is now writing files compliant with NeXus and
the Integrated Spectral Analysis Workbench (ISAW) software from IPNS
is being used with this data. Problems can arise if the standard is
interpreted in different ways and information that belongs in the file
is not accounted for in the standard. This paper will discuss an
inter-laboratory collaboration in relation to a maturing data
standard.
\end{abstract}


\maketitle


\section{Introduction}

Using an international standard file format for data storage is very
appealing. Rather than having to write, or track down, a conversion
utility one can directly use a variety of software packages. For
scattering data the clear choice is to use NeXus.~\cite{nexus;web}
NeXus is a binary file format built on top of the Hierarchical Data
Format (HDF)~\cite{hdf;web} library. The binary format allows for more
efficient data storage while the hierarchical nature of HDF allows for
intelligent grouping of information to describe what is in the file.

The early development of NeXus concentrated on providing a powerful
application program interface (API) implemented in C with wrappers for
FORTRAN 77/90, IDL, and Java. The variety of supported languages
simplifies the work for those who do want to read or write NeXus files
directly. More recently the focus turned to standardizing the
organization of the data and associated information. Originally, the
NeXus standard provided a mechanism for storing information in the
file, but did not spell out a detailed policy for using that
mechanism.  Instrument definitions were incomplete and did not provide
enough information to generate truly portable data files. This was the
main problem when the authors began to collaborate.

\begin{figure}
\begin{center}
  \includegraphics[angle=0,width=3.0in]{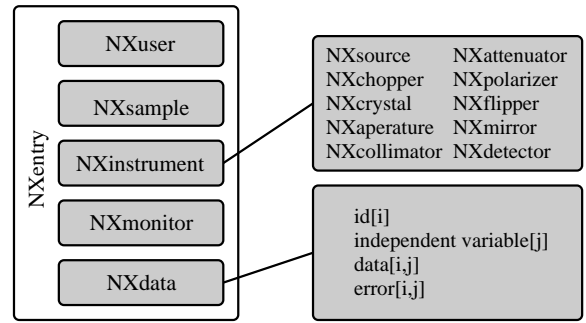}
  \caption{Schematic of the format of a general NXentry. The NXdata
  shown is typical of one dimensional id
  arrays.}\label{fig:fileformat}
\end{center}
\end{figure}

NeXus was starting to define what constituted a generic compliant
file. Each file contains multiple NXentries, and each NXentry is of
the form seen in Fig.~\ref{fig:fileformat}.
The key to NeXus is that while very little information is required, if
it is to be included in the file it has to be included in the manner
specified by the standard.

The Integrated Spectral Analysis Workbench (ISAW) was designed to
visualize and analyze data taken at time-of-flight neutron sources. In
order to appeal to a broader community it looked to support the NeXus
format. The data acquisition system (DAS) team at the Manuel Lujan
Jr. Neutron Scattering Center (MLNSC) was working on an upgrade of
their system which included moving to a more widely used file format
to reduce some of the work involved in analyzing data. Immediately the
two groups tried to share files and found that, while the data could
be written by one system and loaded into the other, much information
was lost due to organizational differences. To solve this problem the
authors decided to specify what needed to be included in a
time-of-flight neutron powder diffractometer (TOFNPD) file. While this
is a conceptually simple instrument, information such as detector
position is needed by many other instruments. This paper will discus
several of the decisions made during a meeting at Los Alamos National
Laboratory concerning the description of detectors.

\section{Identifying Detector Pixels}

\begin{figure}
\begin{center}
  \includegraphics[angle=0,width=2.5in]{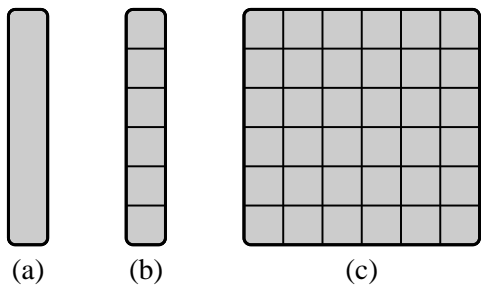}
  \caption{Schematic of an simple tube detector (a), LPSD (b), and
  area detector (c) split into its component
  pixels.}\label{fig:detpix}
\end{center}
\end{figure}

In NeXus, data is stored as a multidimensional rectangular array
inside the NXdata object as seen in Fig.~\ref{fig:fileformat} with
separate arrays for the independent variable and the detector
identifiers. For a simple instrument, with only tube detectors
(Fig.~\ref{fig:detpix}(a)), the data is two dimensional ({\tt
data[i,j]}) with the first axis being time-of-flight ({\tt
time\_of\_flight[i]}) and the second axis being detector number ({\tt
id[j]}). Newer powder diffractometers tend to use linear position
sensitive detectors (LPSD) (Fig.~\ref{fig:detpix}(b)) which would
store data in a three dimensional data array, ({\tt data[i,j,k]}) with
the first axis being the same as the simple tube ({\tt
time\_of\_flight[i]}) and the other axes representing the detector and
pixel number ({\tt id[j,k]}). The case for area detector
(Fig.~\ref{fig:detpix}(c)) is easy to extend from these two cases with
a four dimensional data array ({\tt data[i,j,k,l]}), with the axes
being time-of-flight ({\tt time\_of\_flight[i]}), detector number,
row, and column ({\tt id[j,k,l]}). Since NeXus only provides for
rectangular arrays the case of an instrument with more than one type
of detector must store the data in multiple NXdata. This is also the
case when there is more than one time-of-flight range. To this end the
group decided that if two {\tt data} arrays have the same name
attribute then they are intended to be part of the same data set.

\section{Describing Detectors}

Arguably the most important information, after the data, is a
description of the detector configuration. The detector's position is
needed for conversions to wavelength ($\lambda$), $d$-spacing, and
momentum transfer ($Q$) as well as data corrections such as inelastic
scattering. The solid angle covered is needed to normalize the spectra
for measurement area effects. Other information is useful for
diagnostics such as physical size (length, width, and depth),
orientation, electronics configuration, and type (such as $^3$He or
scintillator). Some information about the detector is redundant as
well. For example if the size and orientation is known then the solid
angle covered is straightforward to calculate and does not need to be
explicitly included. This section will discus some of the properties
of a detector.

\subsection{Specifying Position}

\begin{figure}
\begin{center}
  \includegraphics[angle=0,width=3.0in]{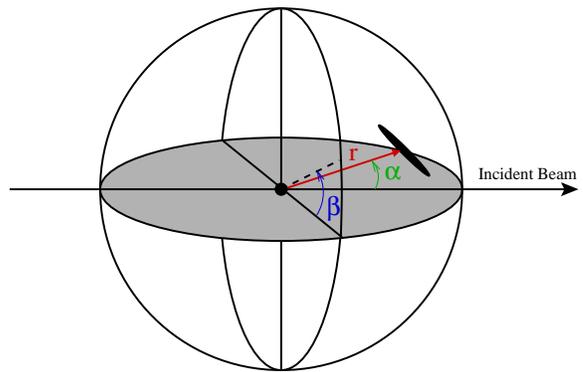}
  \caption{Schematic of a detector at a distance $r$ from the center
  of the instrument. The angles $\alpha$ and $\beta$ are arbitrarily
  labeled and discussed in the text.}\label{fig:detpos}
\end{center}
\end{figure}

Due to the differences in preferred coordinate systems and the variety
of naming conventions, specifying a detector position is more
difficult than one would expect. While the choice of spherical
coordinates is fairly straightforward, the decision of what to call
the angles, and what to reference them from, was a long topic of
discussion.  With diffraction experiments the beam direction is always
special since things such as the Bragg angle~\cite{giaco;bk92} are
with respect to it. Due to this symmetry the preferred coordinate
system is spherical with the pole coinciding with the beam
direction. This is shown in Fig.~\ref{fig:detpos} with the azimuthal
angle being zero at the horizontal plane of the instrument.
The zero point for the azimuthal angle was chosen to be in the
horizontal plane, to coincide with the location of detectors on many
diffractometers and to be consistent with the traditional labeling of
detector position. The angles labeled in the figure are some of the
standard choices used with spherical coordinates. The one way that the
group could speak and not have any inconsistencies was discussing the
angle $\alpha$ as the polar angle and $\beta$ as the azimuthal
angle. Other possibilities for $\alpha$ were Bragg angle, $2\theta$,
$\theta$, and $\phi$. $\beta$ was also possibly $\theta$ or
$\phi$. All greek labels are used in most mathematics texts, (except
$2\theta$) however, their interpretations vary greatly. For this
reason we decided that the most precise way to refer to the angles
were as {\tt polar\_angle} and {\tt azimuthal\_angle}.

\subsection{Specifying Orientation and Size}

\begin{figure}
\begin{center}
  \includegraphics[angle=0,width=3.0in]{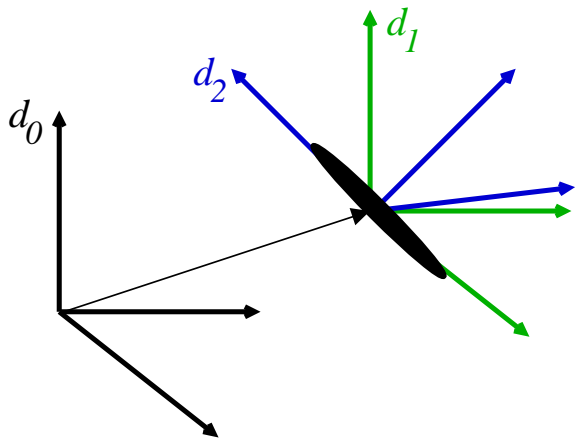}
  \caption{Schematic of specifying the orientation of a detector with
  various possible frames of reference. The labels ($d_0$, $d_1$, and
  $d_2$) are described in the main text.}\label{fig:detorient}
\end{center}
\end{figure}

Another important correction in data analysis is to normalize the
detectors by the solid angle they cover. Therefore, the information
needed is either the solid angle itself, or the physical size (length,
width, and depth) and orientation of the detector so the solid angle
can be calculated. In Fig.~\ref{fig:detorient} a couple of the ways of
determining the detector orientation are shown. 
$d_0$ is the coordinates of the center of the instrument, shown for
clarity. The second frame of reference, $d_1$, is parallel to $d_0$
but shifted to the position of the detector. $d_2$ is coincident with
the major symmetry axes of the detector. Looking at this picture the
simplest description of the orientation is the Euler angles to rotate
$d_1$ to $d_2$.

\section{Summary}

This paper reviewed the results of discussions concerning storing data
in a NeXus file. We described the format for the data itself as well
as specifying the layout of the detectors that measured the data.


\begin{acknowledgments}
Summer support for this work has benefitted through Department of
Educational Programs and the Summer Faculty Research Program. Argonne
National Laboratory is funded by the U.S. Department of Energy,
BES-Materials Science, under Contract W-31-109-ENG-38. Los Alamos
National Laboratory is funded by the US Department of Energy under
contract W-7405-ENG-36.
\end{acknowledgments}


\bibliographystyle{aip}

\end{document}